\documentclass[prl,twocolumn,showpacs,superscriptaddress]{revtex4}
\usepackage{graphicx,bm,amssymb,amsmath}

\newcommand{\Eq}[1]{Eq.~\eqref{#1}}
\newcommand{\eq}[1]{\eqref{#1}}
\newcommand{\Fig}[1]{Fig.~\ref{#1}}

\newcommand{\beq}{\begin{equation}}
\newcommand{\eeq}{\end{equation}}

\newcommand{\beqa}{\begin{eqnarray}}
\newcommand{\eeqa}{\end{eqnarray}}

\newcommand{\Beqa}{\begin{eqnarray*}}
\newcommand{\Eeqa}{\end{eqnarray*}}

\newcommand{\nn}{\nonumber}
\newcommand{\ds}{\displaystyle}

\newcommand{\pdag}{{\phantom{\dagger}}}

\newcommand{\dx}{\partial_x}

\newcommand{\bra}[1]{\left\langle{#1}\right\rvert}
\newcommand{\ket}[1]{\left\lvert{#1}\right\rangle}

\DeclareMathOperator{\sign}{sgn}

\newcommand{\PRL}[3]{Phys. Rev. Lett.~\textbf{#1}, #2 (#3)}
\newcommand{\PRB}[3]{Phys. Rev. B~\textbf{#1}, #2 (#3)}
\newcommand{\JPA}[3]{J. Phys. A~\textbf{#1}, #2 (#3)}
\newcommand{\JPC}[3]{J. Phys. C~\textbf{#1}, #2 (#3)}
\newcommand{\JSM}[2]{J. Stat. Mech.~#1 (#2)}

\newcommand{\etal}{\textit{et al.}}

\begin{document}

\title{
Threshold singularities in the dynamic response of gapless integrable models
}

\author{Vadim V. Cheianov}
\affiliation{Physics Department, Lancaster University, Lancaster, LA1 4YB, UK}

\author{Michael Pustilnik}
\affiliation{School of Physics, Georgia Institute of Technology,
Atlanta, GA 30332, USA}

\begin{abstract}
We develop a method of an asymptotically exact treatment of threshold
singularities in dynamic response functions of gapless integrable models.
The method utilizes the integrability to recast the original problem in terms 
of the low-energy properties of a certain deformed Hamiltonian. The deformed 
Hamiltonian is local, hence it can be analysed using the conventional field theory 
methods.  We apply the technique to spinless fermions on a lattice with 
nearest-neighbors repulsion, and evaluate the exponent characterizing the 
threshold singularity in the dynamic structure factor.
\end{abstract}

\maketitle

One-dimensional (1D) interacting systems~\cite{Giamarchi,KBI,Affleck_review}
occupy a special place in quantum physics. Although interactions have much 
stronger effect in 1D than in higher dimensions, it is often possible to evaluate 
observable quantities exactly. Besides forming the basis of our understanding 
of strong correlations, 1D models have long served as a test bed for various 
approximate methods.

Many properties of interacting 1D systems can be understood by
considering integrable models~\cite{KBI}. The paradigmatic example is
$N$ spinless
fermions on a lattice with $L$ cites with periodic boundary
conditions,
\beq H = \sum_{m=1}^L \left[
-\,\psi_m^\dagger \psi_{m+1}^\pdag - \,\psi_{m+1}^\dagger
\psi_{m}^\pdag + \,2\Delta \,n_m n_{m+1} \,\right].
\label{xxz}
\eeq
Here $n_m = \psi_m^\dagger \psi_{m}^\pdag$ and $\Delta$ is 
the repulsion strength~\cite{eta}. 
The experimentally relevant~\cite{xxz} response function for the
model \eq{xxz} is the dynamic structure factor
\beq
S(q,\omega) = \frac{2\pi}{L}
\sum_f\left|\bra{f} n^\dagger_{q\!} \ket{0}\right|^2
\delta\bigl(\omega - \epsilon_f + \epsilon_0\bigr)
\label{DSF}
\eeq
in the thermodynamic limit $L\to\infty$ taken at a constant filling factor
$\nu=N/L\leq 1/2$. In \Eq{DSF},
$n^\dagger_q = \sum_k \psi^\dagger_k \psi^\pdag_{k-q}$
with $\psi_k = L^{-1/2}\sum_m e^{-i m k}\psi_m$,
$\ket{f}$ is an eigenstate of \Eq{xxz} with energy $\epsilon_f$, 
and $\ket{0}$ is the ground state with $\epsilon_0$ being the ground 
state energy.

\begin{figure}[h]
\includegraphics[width=0.45\columnwidth]{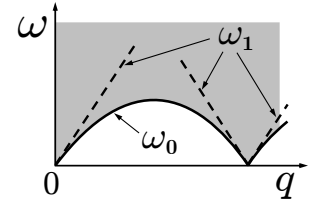}
\caption{
Support of the structure factor in $(\omega,q)$ plane. For $q\neq 0,2k_F$,
the boundary of the support $\omega_0(q)$ lies \textit{below} the straight
dashed lines $\omega_1(q) = \min\bigl\{vq, v|q-2k_F|\bigr\}$; 
here $k_F = \pi\nu$ is the Fermi momentum.
}
\label{fig1}
\end{figure}

In any 1D system, conservation laws restrict the support of correlation 
functions in $(\omega,q)$ plane. For example, $S(q,\omega)>0$ only 
at $\omega > \omega_0(q)$, see \Fig{fig1}. On general 
grounds, $S(q,\omega)$ is expected to exhibit a 
power-law singularity at the threshold~\cite{struc_fac},
\beq
S(q,\omega) \propto \bigl[\omega - \omega_0(q)\bigr]^{-\mu}.
\label{threshold}
\eeq

Although exact eigenstates of integrable models can be constructed 
using the Bethe ansatz~\cite{KBI}, evaluation of dynamic correlation 
functions is very difficult. A considerable progress was achieved in 
understanding \textit{gapful} models~\cite{EK}. However, it is still 
largely an open problem in the \textit{gapless} case, and, with few 
exceptions (see, e.g.,~\cite{Haldane-Shastry,Calogero,CZG,SP}),
the threshold exponents $\mu$ are not known.
For the model \eq{xxz}, the most complete results so far were obtained 
by combining numerics with the algebraic Bethe ansatz~\cite{Caux,Pereira}.
The main limitation of this technique is very slow convergence towards 
the thermodynamic limit, which makes it very difficult to evaluate the exponent.

In the alternative Luttinger liquid approach~\cite{Giamarchi,Affleck_review,Haldane},
a 1D system is described by two parameters, the sound velocity 
$v = (d\omega_0/dq)_{q\to 0}$, and the Luttinger parameter $\kappa$
characterizing the interaction strength.
At $q\to 2k_F$, Luttinger liquid theory yields \Eq{threshold} with the exponent
\beqa
\mu_L = 1-\kappa.
\label{mu_L}
\eeqa

At $q \to 0$, however, one finds $S\propto \delta(\omega - vq)$. 
The discrepancy with \Eq{threshold} is due to the omission in the fixed-point
Luttinger liquid Hamiltonian~\cite{Affleck_review,Haldane} of the irrelevant
in the RG sense operators~\cite{Pereira,Haldane} representing the spectrum
nonlinearity. Indeed, for small $q$, most of the spectral weight of $S(q,\omega)$
is confined to a narrow interval of $\omega$ of the width
$\delta\omega\sim \omega_1-\omega_0$ about $\omega=\omega_1$,
while \Eq{threshold} is applicable at
$\omega-\omega_0\ll\delta\omega$~\cite{struc_fac}.
For a linear spectrum $\delta\omega=0$, which makes the regime of \Eq{threshold}
inaccessible. For the same reason, the Luttinger liquid result \eq{mu_L}
is valid, strictly speaking,  only at $\omega-\omega_0 \gg \delta\omega$,
and the exact exponent may differ from $\mu_L$.

In fact, the true threshold exponents often deviate from the Luttinger liquid
theory predictions already in the lowest order in the interaction
strength~\cite{Khodas}; for $S(q,\omega)$ near $q=2k_F$, such deviations
show up at $q>2k_F$.

In this Letter we develop a technique that allows exact evaluation of
the exponents characterizing threshold singularities. The technique is 
applicable to any 
correlation function that exhibits a threshold behavior and to any
gapless model integrable by the Bethe ansatz. We illustrate the idea of the
method by working out the dynamic structure factor \eq{DSF} for the
model \eq{xxz} as an example.

Among various states $\ket{f}$ for which
$\bra{f} n^\dagger_{q\!} \ket{0}\neq 0$
[see \Eq{DSF}] for a given $q$, one state, say, $\ket{f_q}$, has the lowest energy
$\epsilon_q = \epsilon_0 + \omega_0(q)$. Consider now a \textit{deformed}
Hamiltonian $\widetilde H$ with the following properties:
\begin{itemize}
\item[(i)]
The deformed Hamiltonian $\widetilde H$ is local.
\item[(ii)]
$\widetilde H$ commutes with $H$, so that $H$ and $\widetilde H$ have 
the same set of
eigenstates $\ket{f}$.
\item[(iii)]
The states $\ket{0}$ and $\ket{f_q}$ represent the
doubly-degenerate ground state of $\widetilde H$.
\item[(iv)]
The deformation $H\to \widetilde H$ is \textit{continuous} in the sense
that if the state $\ket{f}$ has momentum $q$ and its energy $\epsilon_f$
is close to $\epsilon_q$, then the corresponding eigenvalues 
of $\widetilde H$ satisfy
$\widetilde\epsilon_f - \widetilde\epsilon_q =
\epsilon_f - \epsilon_q + O\bigl[(\epsilon_f - \epsilon_q)^2\bigr]$;
similarly, 
$\widetilde\epsilon_f - \widetilde\epsilon_0 =
\epsilon_f - \epsilon_0 + O\bigl[(\epsilon_f - \epsilon_0)^2\bigr]$
for states $\ket{f}$ with zero momentum.
\end{itemize}

Once the Hamiltonian $\widetilde H$ satisfying these conditions has 
been constructed, it can be analysed using conventional field-theoretical 
methods~\cite{Affleck_review}. In particular, condition (i) allows one 
to identify the low-energy projections of microscopic fields with local 
operators in the effective continuum description. The coupling constants 
of this effective low-energy theory can be found by comparing its low-energy 
spectrum with that of $\widetilde H$. (Essentially the same ideas are behind 
the Luttinger liquid description of the low-energy excitations of spin
chains~\cite{Affleck_review}). Finally, conditions (ii)-(iv) guarantee that
the structure factor calculated for the Hamiltonian $\widetilde H$ will 
have a power-law singularity at $\omega\to 0$ with the same exponent 
$\mu$ that characterizes the threshold behavior in the original model.

Integrable models have an infinite number of independent
\textit{local} operators $I_n$ commuting with $H$ (integrals of
motion). Thus, any Hamiltonian of the form
\beq 
\widetilde H =
\sum c_n I_n 
\label{H_q}
\eeq
will satisfy conditions (i) and (ii). 

\begin{figure}[h]
\includegraphics[width=0.65\columnwidth]{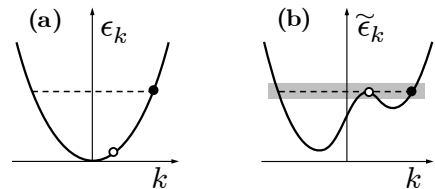}
\caption{
(a) For free fermions $(\Delta=0)$ and for $\nu<1/2$ and $q<2k_F$, the state
$\ket{f_q}$ consists of a single particle-hole pair added to the Fermi sea:
the particle is created at the Fermi momentum $k=k_F$, and the hole
at $k = k_F-q$.
(b) Single-particle spectrum of the deformed Hamiltonian $\widetilde H$:
only states in a narrow strip of energies (shaded) about the Fermi level (dashed line)
contribute to the structure factor near the threshold.
}
\label{fig2}
\end{figure}

For free fermions
$\ket{f_q}= \psi^\dagger_{k_F} \psi^\pdag_{k_F-q}\ket{0}$,
see \Fig{fig2}\,(a), 
and the integrals of motion have a simple form. 
Conditions (ii)-(iv) will be fulfilled if the single-particle spectrum 
of the deformed Hamiltonian has the shape sketched in \Fig{fig2}\,(b).
While this is not sufficient to determine the coefficients $c_n$ uniquely, 
the low-energy spectrum of $\widetilde H$ is completely specified.

In the Bethe ansatz, excitations of integrable models such as \Eq{xxz} 
are classified in terms of fermion-like quasiparticles and quasiholes~\cite{KBI}.
Similar to free fermions, the state $\ket{f_q}$ corresponds to a
quasiparticle with momentum $k_F$ and a quasihole with momentum 
$k_F-q$ added to the ground state, cf. \Fig{fig2}\,(a). 

Provided that the quasiparticle spectrum of the deformed Hamiltonian 
(see below) is similar to that shown in \Fig{fig2}\,(b), 
it is obvious that the infrared fixed point of $\widetilde H$ corresponds to 
a single hole minimally coupled to the right- and left-moving 
fermions with linear spectrum. 
We introduce bosonic fields $\varphi_{\pm}$ which satisfy
$\bigl[\varphi_\alpha(x),\varphi_{\alpha'}(y)\bigr]
= i\pi\alpha\,\delta_{\alpha,\alpha'}\sign(x-y)$,
and the field $d$ which describes an infinitely heavy hole~\cite{heavy_hole}. 
The fixed-point Hamiltonian then assumes the form familiar from the x-ray 
edge singularity problem~\cite{Balents},
\beq
\widetilde H = \int\frac{dx}{4\pi}\sum_{\alpha} v_\alpha\Bigl[
 (\dx\varphi_\alpha)^2
- 2\beta_\alpha (\dx\varphi_{\alpha\!})\,d(x) d^\dagger(x) \Bigr].
\label{fixed-point}
\eeq

The low-energy projection of the microscopic field $\psi_m$ is given by
$\psi_m=\sum_{\alpha}e^{i \alpha k_F x} \psi_\alpha(x) +\, e^{ i (k_F-q) x} d(x)$,
where $m=x$ is treated as a continuous variable and the fields $\psi_{\pm}$
are related to $\varphi_\pm$ according to
\beq
\psi_\alpha \propto
\exp\bigl[i\alpha\bigl(\varphi_\alpha \cosh\vartheta -
\varphi_{-\alpha}\sinh\vartheta\bigr)\bigr], 
\quad 
e^{-\,2\vartheta} = \kappa\,.
\label{bosonization}
\eeq
The leading contribution to the density operator
$n^\dagger_q$ [see \Eq{DSF}] is then given by
\beq
n_q^\dagger \propto \int\!dx\,\psi^\dagger_{+}(x) d(x).
\label{density}
\eeq
Evaluation of the structure factor \eq{DSF} using Eqs. \eq{fixed-point}-\eq{density}
yields a power-law singularity with the exponent
\beq
\mu = 1 - \left(\cosh\vartheta + \frac{\,\beta_{+}}{2\pi} \right)^2
- \left(\sinh\vartheta + \frac{\,\beta_{-}}{2\pi} \right)^2.
\label{exponent}
\eeq

We now sketch the construction of the deformed Hamiltonian~\eq{H_q} 
and the derivation of the coupling constants of the corresponding 
fixed-point Hamiltonian~\eq{fixed-point}.
We choose $I_0 = \sum_m \psi^\dagger_m \psi^\pdag_m$,
so that $c_0$ in \Eq{H_q} plays the role of the chemical potential.
For $n>0$, the integrals of motion are expressed via the derivatives of the 
transfer matrix (trace of the monodromy matrix) 
$\tau(\xi)$ with respect to the spectral parameter $\xi$~\cite{KBI},  
\beq
I_{n>0}= i\bigl[\,d^n\!\ln\tau/d\xi^n\bigr]_{\,\xi \to i\pi/2 - i \eta}.
\eeq
The first operator in this hierarchy is proportional to the Hamiltonian 
itself: $I_1= H/\sin (2\eta)$. The next one, $I_2$, is 
given in a closed form in~\cite{I_2}.

Consider a quasiparticle (quasihole) excitation of the Hamiltonian~\eq{xxz} 
characterized by the rapidity $\lambda$~\cite{rapidity}. By construction, 
such excitation is an eigenstate of the deformed Hamiltonian $\widetilde H$, 
see~\Eq{H_q}. Using properties of the transfer matrix~\cite{KBI}, 
one can show~\cite{unpub} that the corresponding eigenvalue 
$\widetilde\epsilon(\lambda)$ satisfies the equation
\beq
\widetilde \epsilon(\lambda) -
\frac{1}{2\pi}\!\int_{-\lambda_F}^{\lambda_F}\!\!d\mu\, 
K(\lambda-\mu)\,\widetilde\epsilon(\mu)
= c_0+ \sum_{n>0} c_n (-1)^n \,\frac{d^n p_0}{d\lambda^n }\,.
\label{tile}
\eeq
Here the Fermi rapidity $\lambda_F$ is the solution to 
$k(\lambda_F) = k_F$~\cite{rapidity},
$K(\lambda)=d\theta/d\lambda$ where $\theta(\lambda)$ 
is the bare two-particle phase shift, and $p_0(\lambda)$ is the bare 
particle momentum; for the model~\eq{xxz} these are given by~\cite{KBI}
\beq
\theta= i \ln\!\left[\frac{\sinh(\lambda+2i\eta)}{\sinh(-\lambda+2i \eta)}\right],
\quad
p_0 = i\ln\!
\left[\frac{\cosh(\lambda-i\eta)}{\cosh(\lambda+i\eta)}\right].
\label{theta}
\eeq

In order to satisfy the conditions (ii)-(iv) above,  
we impose additional constraints on $\widetilde\epsilon(\lambda)$,
\beqa
& 
\widetilde\epsilon(\lambda_q) = \widetilde\epsilon(\pm\,\lambda_F) = 0, 
\quad
(d\widetilde\epsilon/d\lambda)_{\lambda_q} = 0,
\label{constraints} 
\\
&\ds
(d\widetilde\epsilon/d\lambda)_{\pm\,\lambda_F}
= (d\epsilon/d\lambda)_{\pm\,\lambda_F}
- \frac{\rho(\lambda_F)}{\rho(\lambda_q)}\,(d\epsilon/d\lambda)_{\lambda_q},
\nn
\eeqa
where $\lambda_q$ is the solution to 
$k(\lambda_q) = k_F-q$~\cite{rapidity}.
The constraints are equivalent to five \textit{linear} 
equations on the coefficients $c_n$ in Eqs. \eq{H_q} and \eq{tile}. 
In order to satisfy these equations, it is sufficient to keep the first five 
integrals of motion in \Eq{H_q}. 

The coupling constants of the effective fixed-point Hamiltonian
\eq{fixed-point} follow from the comparison of the finite-size
spectrum of \eq{fixed-point} with the exact low-energy spectrum 
of the deformed Hamiltonian $\widetilde H$.
This procedure is standard~\cite{KBI} and yields~\cite{unpub}
\beq 
\beta_\pm= \pm\,2\pi F(\pm \,\lambda_F\vert\lambda_q), 
\label{mainanswer}
\eeq
where $F$ is the dressed phase shift that satisfies~\cite{KBI}
\beq
F(\lambda\vert\zeta)
- \frac{1}{2\pi}\!\int_{-\lambda_F}^{\lambda_F}\!\!d\mu\, 
K(\lambda-\mu)F(\mu \vert\zeta) 
= \frac{1}{2\pi}\,\theta(\lambda-\zeta). 
\label{dressedphase}
\eeq
Eqs. \eq{mainanswer} and \eq{dressedphase} uniquely define the 
parameters $\beta_{\pm}$, and, therefore, the exponent 
\eq{exponent}.

Note that $\beta_\pm$ do not depend explicitely on 
the coefficients $c_n$. Indeed, the constraints \eq{constraints} 
do not completely fix $\widetilde H$ but only specify its low-energy 
spectrum. We emphasize that our construction does not rely on the 
model-specific \Eq{theta} but is applicable to any model integrable 
by the algebraic Bethe ansatz.

We now use Eqs. \eq{exponent}, \eq{mainanswer}, and \eq{dressedphase} 
to evaluate the threshold exponent for the model \eq{xxz}. 
Precisely at half-filling $\lambda_F\to \infty$~\cite{KBI}
and \Eq{dressedphase} is solved by Fourier transform with the result
\beq
F(\pm \,\infty\,\vert\zeta)= \pm \, (\kappa-1)/2\,, 
\label{14}
\eeq
where we used the well-known~\cite{KBI} value of the Luttinger
parameter at half-filling, $\kappa = \pi/4\eta$~\cite{eta}.
Eqs.~\eq{exponent}, \eq{mainanswer}, and \eq{14} then yield
a momentum-independent exponent
\beq
\mu_0 =
\frac{1-\kappa}{2}\left(\kappa-\frac{1}{\kappa} +
\frac{2}{\sqrt{\kappa}}\right),
\quad
\nu=1/2\,.
\label{mu_0}
\eeq
Comparison with \Eq{mu_L} shows that the exact exponent $\mu_0$ is
\textit{smaller} than the Luttinger liquid result $\mu_L$ (note that
$\kappa$ varies between $1/2$ and $1$). For a weak interaction,
$\mu_L - \mu_0 \approx \mu_L^2/2\approx 2(\Delta/\pi)^2$.

\begin{figure}[h]
\includegraphics[width=0.75\columnwidth]{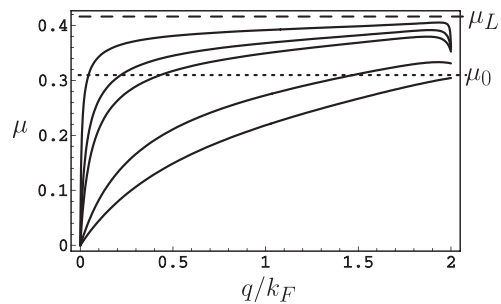}
\caption{
Threshold exponent $\mu(q)$
for $\Delta = 0.9$ and for $\nu = 0.4,0.45,0.49,0.495,0.499$ (bottom to top).
The dashed horizontal lines correspond to $\mu_L$ and $\mu_0$ at half-filling.
}
\label{fig3}
\end{figure}

Away from half-filling, \Eq{dressedphase} can be solved numerically. 
The resulting exponent is $q$-dependent, see \Fig{fig3}. It varies from 
$\mu(0)=0$ to $\mu(2k_F) =\mu_2$ with $\mu_0<\mu_2<\mu_L$;
the exact value of $\mu_2$ depends on both $\Delta$ and $\nu$~\cite{unpub}.
Very close to half-filling, the dependence $\mu(q)$ is nonmonotonic:
outside narrow intervals of the width
\beq
\delta k \sim 1/2-\nu\ll k_F
\label{16}
\eeq
near $q=0$ and $q=2k_F$, the exponent approaches a constant,
$\mu(q)\approx\mu_1=\text{const}$.
Surprisingly, $\mu_1$ coincides with the Luttinger
liquid exponent $\mu_L$ rather than with the exact half-filling result $\mu_0$.

This discrepancy originates in the peculiar behavior of the phase shifts
near the Fermi points. Consider \Eq{dressedphase} at
$|\zeta\pm\lambda_F|\gg \delta\lambda$ with $\delta\lambda=1-2\eta/\pi$
(this limit corresponds to $q$ away from $q=0,2k_F$).
In order to find the phase shift $F(\lambda\vert\zeta)$ for 
$\lambda_F\gg \delta\lambda$ (i.e., close to half-filling) and 
$\lambda\approx\lambda_F$ (i.e., close to the right Fermi point), 
we replace $\theta(\lambda-\zeta)$ in the r.h.s of \eq{dressedphase} 
by $\theta(\infty)$, and extend the integration in the l.h.s. to $-\infty$.
The resulting equation
\beq
F(\lambda\vert\zeta)- \frac{1}{2\pi} \int_{-\infty}^{\lambda_F}\!d\mu\,
K(\lambda-\mu)F(\mu \vert\zeta) = \frac{1}{2\pi}\,\theta(\infty),
\label{17}
\eeq
as well as the similar equation for $\lambda\approx-\lambda_F$,
describes the fractional charge function~\cite{KBI}. Its solution yields
\begin{equation}
2\pi F(\lambda\vert\zeta) = \left\{
\begin{array}{lc}
\gamma_0, & \lambda, \lambda_F - \lambda\gg\delta\lambda
\\
\gamma_1, & \lambda_F - \lambda\ll \delta\lambda
\end{array}
\right.
\label{18}
\end{equation}
with $\gamma_0 = \pi(\kappa-1)$ and $\gamma_1 =
\pi(\kappa-1)\kappa^{-1/2}$. In other words, the limits
$\lambda\to\pm\,\lambda_F$ and $\lambda_F\to\infty$ for the phase shift
$F(\lambda\vert\zeta)$ do not commute. If the limit $\lambda\to
\pm\,\lambda_F$ is taken first, and the resulting phase shifts are
substituted into \Eq{mainanswer}, one finds $\beta_\pm = \gamma_1$.
\Eq{exponent} then yields $\mu_1=\mu_L$. However, by taking first
the limit $\lambda_F\to\infty$ (i.e., $\nu\to 1/2$), one would find
$\beta_\pm = \gamma_0$ and $\mu_1 = \mu_0$.

The noncommutativity of limits has observable consequences. 
Indeed, $F(\lambda\vert\zeta)$ characterizes the strength
of the interaction between a quasihole at rapidity $\zeta$ and a quasiparticle
at rapidity $\lambda$. According to \Eq{18}, the phase shift
at $\lambda\approx\pm\,\lambda_F$ changes with $\lambda$ on the scale
$\delta\lambda\ll\lambda_F$. In the momentum space, this corresponds
to narrow intervals of the width
$\delta k = 2\pi \rho(\lambda_F)\delta\lambda\sim(1/2-\nu)$
near $k=\pm\, k_F$ (here we used the well-known~\cite{KBI}
result for $\rho(\lambda)$ near half-filling).
States within or outside these intervals interact with the quasihole with
coupling constants $\beta_\pm \approx \gamma_1$ or 
$\beta_\pm \approx \gamma_0$, respectively.

As the filling factor approaches $1/2$, the interval $\delta k$ collapses.
For a finite-size system close to half-filling, $\delta k$ will eventually
become compatible with the momentum quantum $\sim 1/L$.
In this limit, the threshold behavior is dominated by states outside the
interval $\delta k$. Accordingly, the exponent $\mu_1$ that characterizes
the threshold singularity away from $q=0,2k_F$ exhibits a crossover
from $\mu_1= \mu_L$ at $1\gg\delta k\gg 1/L$ to $\mu_1= \mu_0$ at
$\delta k\ll 1/L$.

In the recent study~\cite{PWA} the exponent $\mu$ at half-filling
was found to be equal to $\mu_L$. Our consideration shows that
$\mu$ indeed approaches this value when $\nu\to 1/2$. However,
because the limits $\nu\to 1/2$ and $\omega\to\omega_0$ do not
commute, the region of applicability of the result $\mu=\mu_L$
is limited to $\omega-\omega_0\ll v\delta k$. Precisely at half-filling
$\delta k=0$, and the exponent is given by \Eq{mu_0} instead.

It should be mentioned that the two-spinon contribution to the structure
factor has a square-root singularity at $\omega\to\omega_0$~\cite{2-spinon}.
This result was obtained by approaching $\Delta=1$ from the 
\textit{gapful} side of the transition $\Delta>1$. We found $\mu_0<1/2$ 
in the \textit{gapless} regime $\Delta<1$, see  \Eq{mu_0} above.
The discrepancy suggests that the threshold exponent $\mu_0(\Delta)$ 
has a discontinuity at $\Delta=1$.

Finally, for $\delta k>0$ and when $q$ approaches either $0$ or $2 k_F$,
the situation is complicated by the competition between two 
small energy scales, $v\delta k$ and $\delta\omega$ (see above). 
The behavior of the structure factor in this regime will be discussed in 
details elsewhere~\cite{unpub}. At small $q$, it agrees with the first-order 
result of~\cite{struc_fac}:
$\mu(q) \sim (\Delta/\delta k)\,q$ for $0<\delta k\ll 1$ and
$\mu_0\approx\mu_L \sim \Delta$ for $\delta k = 0$; the two
exponents merge at $q \sim \delta k$.

To conclude, we proposed a method of evaluating the exponents
characterizing threshold singularities in the dynamic response functions 
of gapless integrable models.  Application of the method to the dynamic 
structure factor of 1D spinless fermions on a lattice revealed unexpected 
complexity in the dependence of the threshold exponent on the system 
parameters near half-filling.

\begin{acknowledgments}
We thank I. Affleck, J.-S. Caux, R. Pereira, and M. Zvonarev for discussions.
Research at Georgia Tech is supported by DOE grant DE-FG02-ER46311.
\end{acknowledgments}


\begin{thebibliography}{99}
\bibitem{eta}
In the gapless regime $0<\Delta<1$, and it is convenient to parametrize
$\Delta$ as $\Delta=-\cos2\eta$ with $\pi/4<\eta<\pi/2$.
\bibitem{xxz} The structure factor \Eq{DSF} describes the optical
absorption rate of the system. Alternatively, the model \eq{xxz} can
be mapped onto $S=1/2$ antiferromagnetic spin chain, see,
e.g.,~\cite{Affleck_review}. \Eq{DSF} then gives the spin structure
factor measured by inelastic neutron scattering.
\bibitem{heavy_hole}
Keeping the hole mass finite introduces an irrelevant in the RG 
sense perturbation that does not affect the value of the threshold exponent~\cite{Balents}.
\bibitem{rapidity}
The physical momentum $k$ corresponding to the rapidity 
$\lambda$ is given by 
$k(\lambda) = 2\pi\!\int_0^\lambda\!d\lambda\rho(\lambda)$,
where the density of rapidities in the thermodynamic limit $\rho(\lambda)$ 
satisfies the equation~\cite{KBI} 
$2\pi\rho(\lambda) - \int_{-\lambda_F}^{\lambda_F}\!\!d\mu\,
K(\lambda-\mu) \rho(\mu) = d p_0/d\lambda$\,.
\end{thebibliography}
\end{document}